\documentclass{aipproc}

\layoutstyle{6x9}

\usepackage{amssymb,amsmath}

\newcommand{\eq}[1]{\begin{equation} #1 \end{equation}}

\begin{document}

\title{On the pulsating strings in Sasaki-Einstein spaces}


\author{D.~Arnaudov}{address={Department of Physics, Sofia University, 5 J. Bourchier Blvd, 1164 Sofia, Bulgaria}}

\author{H.~Dimov}{address={Department of Physics, Sofia University, 5 J. Bourchier Blvd, 1164 Sofia, Bulgaria}}

\author{R.C.~Rashkov}{address={Institute for Theoretical Physics, Vienna University of Technology, Wiedner Hauptstr. 8-10, 1040 Vienna, Austria},altaddress={Department of Physics, Sofia University, 5 J. Bourchier Blvd, 1164 Sofia, Bulgaria}}

\begin{abstract}
We study the class of pulsating strings in $AdS_5\times Y^{p,q}$ and $AdS_5\times L^{p,q,r}$. Using a generalized ansatz for pulsating string configurations, we find new solutions for this class in terms of Heun functions, and derive the particular case of $AdS_5\times T^{1,1}$, which was analyzed in arXiv:1006.1539~[hep-th]. Unfortunately, Heun functions are still little studied, and we are not able to quantize the theory quasi-classically and obtain the first corrections to the energy. The latter, due to AdS/CFT correspondence, is supposed to give the anomalous dimensions of operators of the gauge theory dual ${\cal N}=1$ superconformal field theory.
\end{abstract}

\maketitle


\section{Introduction}
The attempts to establish a correspondence between the large N
limit of gauge theories and string theory has more than 30 years history and
 over the years it showed different faces. Recently an explicit realization of
this correspondence was provided by the Maldacena conjecture about AdS/CFT correspondence
\cite{Maldacena}. The convincing results from the duality between type IIB string
theory on $AdS_5\times S^5$ and ${\cal N}=4$ super Yang-Mills theory
\cite{Maldacena,GKP,Witten} made this subject  a major research area, and many
fascinating new features have been established.

After the impressive achievements in the most supersymmetric example of AdS/CFT correspondence,
namely $AdS_5\times S^5$, it is important to extend the considerations to less supersymmetric
gauge theories, moreover that the latter are more interesting from physical point of view. The integrability is expected to play the same crucial role, but unfortunately not much is known on the subject. One way to find a theory with less supersymmetry is to take a stack of N D3 branes and place them at the apex of a conifold \cite{Klebanov:1998hh}. This model possesses a lot of interesting features and allows to build gauge theory operators of great physical importance. The resulting ten dimensional spacetime is the direct product $AdS_5\times T^{1,1}$. Since then, it was found that for five-dimensional Sasaki-Einstein manifolds there is an infinite family of inhomogeneous metrics on $Y^{p,q}\cong S^2\times S^3$, which is characterized by relatively prime positive integers $p,\,q\,$ with $\,0<q<p\,$ \cite{Gauntlett:2004zh,Gauntlett:2004yd,Gauntlett:2006}. In this case, there is an effective action of a torus $T^3\cong U(1)^3$ on the Calabi-Yau cone of $Y^{p,q}$, which preserves the symplectic form and metric on it since it is an isometry. In~\cite{Martelli:2004wu} there is an extensive discussion on the geometric features of these manifolds, and the superconformal quiver gauge theories dual to type IIB string theory on $AdS_5 \times Y^{p,q}$ were proposed in \cite{Benvenuti:2004dy}-\cite{Butti:2005sw}.

The  spaces $\,Y^{p,q}$ are of cohomogeneity one, but the correspondence can be generalized to spaces with cohomogeneity two, called $\,L^{p,q,r}$ spaces \cite{Cvetic:2005ft}. These are characterized by the relative positive coprime integers $p,\,q$ and $r$ with $0<p\leq q,\,0<r<p+q$, and with $p,\,q$ to be coprime to $s=p+q-r$, and have isometry $U(1)\times U(1)\times U(1)$. The metrics $Y^{p,q}$ are a special case of $L^{p,q,r}$ where $p+q=2r$. Further developments
can be traced in~\cite{Gauntlett:2004zh}-\cite{Giataganas:2009}.

The semi-classical strings have played, and still play, an important role in studying various aspects of
$AdS_5/SYM_4$ correspondence \cite{Bena:2003wd}-\cite{Lee:2008sk}. The development in this subject gives a strong hint about how the new emergent duality can be investigated. An important class of semi-classical string solutions is the class of pulsating strings introduced first in \cite{Minahan:2002rc}, and studied and generalized further in \cite{Engquist:2003rn,Dimov:2004xi,Smedback:1998yn}. In the case of Sasaki-Einstein backgrounds the pulsating strings are also expected to play an essential role, and thorough analysis and quasi-classical quantization was provided in \cite{Arnaudov:2010} for $T^{1,1}$. The purpose of this paper is to throw some light on the class of pulsating strings in $Y^{p,q}$ and $L^{p,q,r}$.

The paper is organized as follows. In the next section we give the pulsating strings and their quasi-classical quantization for the cases of $AdS_5\times Y^{p,q}$ and $AdS_5\times L^{p,q,r}$ backgrounds, restricting the string dynamics to $Y^{p,q}$ and $L^{p,q,r}$ parts of the spacetime. The third section is devoted to the derivation of the wave functions associated with the Laplace-Beltrami operators on $Y^{p,q}$ and $L^{p,q,r}$. Also, we derive the $T^{1,1}$ case from $Y^{p,q}$. We conclude with a brief discussion on the results.

\section{Pulsating strings in $AdS_5\times Y^{p,q}$ and $AdS_5\times L^{p,q,r}$}
We consider a circular pulsating string expanding and contracting only in the $Y^{p,q}$ part of $AdS_5\times Y^{p,q}$ ($R\times Y^{p,q}$). Then, the relevant metric we will work with is given by
\begin{equation}
ds^2_{t\times Y^{p,q}}=R^2\left(-dt^2+ds^2_{Y^{p,q}}\right),
\end{equation}
where the metric of $Y^{p,q}$ is
\eq{
\begin{split}
ds^2_{Y^{p,q}}&=\frac{(1-cy)d\theta^2}{6}+\frac{dy^2}{w(y)q(y)}+\left(\frac{(1-cy)\sin^2\theta}{6}+\Omega(y)\cos^2\theta\right)d\phi^2+w(y)d\alpha^2+\\
&\Omega(y)d\psi^2-2w(y)f(y)\cos\theta d\phi d\alpha-2\Omega(y)\cos\theta d\phi d\psi+2w(y)f(y)d\alpha d\psi,
\end{split}
\label{metric_Ypq}
}
where
\begin{equation}
\begin{split}
\Omega(y)&=\frac{q(y)}{9}+w(y)f(y)^2,\\
w(y)&=\frac{2(a-y^2)}{1-cy},\,q(y)=\frac{a-3y^2+2cy^3}{a-y^2},\,f(y)=\frac{ac-2y+cy^2}{6(a-y^2)}.
\end{split}
\end{equation}
In addition, the three real roots of $q(y)$ are chosen so that $y_1<0<y_2<y_3$. The coordinates have the ranges
\eq{
0\leq\theta\leq\pi,\,y_1\leq y\leq y_2,\,0\leq\phi<2\pi,\,0\leq\psi\leq2\pi.
}
Having in mind the explicit form of the $Y^{p,q}$ metric \eqref{metric_Ypq}, one can write it as
\begin{equation}
ds^2_{Y^{p,q}}=G_{ij}dx^i dx^j+\hat{G}_{pq}dy^p dy^q,
\end{equation}
where the part $G_{ij}$ is defined by
\begin{equation}
G_{ij}=diag\left( \frac{1-cy}{6},\, \,\frac{1}{w(y)q(y)}\right),\,\,\,\,i,j=1,\,2,\,\,\,\,
x^1=\theta,\,\,x^2=y,
\end{equation}
while $\hat{G}_{pq}=\hat{G}_{pq}(\theta,\,y)$ is the remaining part of metric associated with
$\phi,\,\alpha,\,\psi$ coordinates, denoted here as $p,\,q=1,\,2,\,3,\,\,\,\,y^1=\phi,\,\,y^2=\alpha,\,\,y^3=\psi$
\begin{equation}
(\hat{G}_{pq})=
\left(
\begin{array}{ccc}
\left(\frac{(1-cy)\sin^2\theta}{6}+\Omega(y)\cos^2\theta\right) & -w(y)f(y)\cos\theta & -\Omega(y)\cos\theta \\
-w(y)f(y)\cos\theta & w(y) & w(y)f(y) \\
-\Omega(y)\cos\theta & w(y)f(y) & \Omega(y)
\end{array}
\right).
\end{equation}

The residual worldsheet symmetry allows to identify $t$ with $\tau$, and to obtain a classical pulsating string solution we use the following ansatz
\begin{align}
x^1&=x^1(\tau)=\theta(\tau),\,\,\,\, &x^2=x^2(\tau)=y(\tau),&\,\,\,\,\\
y^1&=\phi=m_1\sigma+h^1(\tau),&y^2=\alpha=m_2\sigma+h^2(\tau),\quad\,\,\,\,&y^3=\psi=m_3\sigma+h^3(\tau).
\end{align}

We are interested in the induced worldsheet metric, which in our case has the form
\begin{equation}
ds^2_{ws}=R^2\left\lbrace\left(-1+ G_{ij}\dot{x}^i \dot{x}^j+\hat{G}_{pq}\dot{h}^p \dot{h}^q\right)d\tau^2 +(\hat{G}_{pq}m_p m_q)d\sigma^2+
2(\hat{G}_{pq}m_p \dot{h}^q)d\tau d\sigma \right\rbrace.
\end{equation}
The Nambu-Goto action in this ansatz reduces to the expression
\begin{equation}
S_{NG}=-TR^2\int d\tau d\sigma
\sqrt{\left(1- G_{ij}\dot{x}^i \dot{x}^j-\hat{G}_{pq}\dot{h}^p \dot{h}^q\right)(\hat{G}_{pq}m_p m_q)+
(\hat{G}_{pq}m_p \dot{h}^q)(\hat{G}_{pq}m_q \dot{h}^p)},
\end{equation}
where $TR^2=\sqrt{\lambda}$.
For our considerations it is useful to pass to Hamiltonian
formulation. For this purpose, we have to find first the canonical momenta of our dynamical system. Straightforward calculations give for the
momenta
\begin{equation}
\Pi_i=\sqrt{\lambda}\dfrac{(\hat{G}_{pq}m_p m_q)G_{ij} \dot{x}^j}{\sqrt{\left(1- G_{ij}\dot{x}^i \dot{x}^j-\hat{G}_{pq}\dot{h}^p \dot{h}^q\right)(\hat{G}_{pq}m_p m_q)+
(\hat{g}_{pq}m_p \dot{h}^q)(\hat{G}_{pq}m_q \dot{h}^p)}},\,\,\,\,i=1,2,
\end{equation}
\begin{equation}
\hat{\Pi}_p=\sqrt{\lambda}\dfrac{(\hat{G}_{pq}m_p m_q)\hat{G}_{pq}\dot{h}^q-
(\hat{G}_{pq}m_p \dot{h}^q)\hat{G}_{pq}m_q}
{\sqrt{\left(1- G_{ij}\dot{x}^i \dot{x}^j-\hat{G}_{pq}\dot{h}^p \dot{h}^q\right)(\hat{G}_{pq}m_p m_q)+
(\hat{G}_{pq}m_p \dot{h}^q)(\hat{G}_{pq}m_q \dot{h}^p)}},\,\,\,\,p=1,2,3.
\end{equation}
Solving for the derivatives in terms of the canonical momenta and substituting back into the Legendre transform of the Lagrangian, we find the Hamiltonian
\begin{equation}
H^2= G^{ij}\Pi_i \Pi_j+\hat{G}^{pq}\hat{\Pi}_p\hat{\Pi}_q+\lambda\,(\hat{G}_{pq}m_p m_q).
\end{equation}
The interpretation of this relation is as in the case of $AdS_5\times S^5$. Namely, the first term in the brackets is the kinetic term while the second one is considered as a potential $V$. The approximation where our considerations are valid assumes high energies, which suggests that one can think of this potential term as a perturbation. For later use we write down its explicit form
\begin{multline}
V(\theta,y)=\lambda\,\hat{G}_{pq}(\theta,\,y)m_p m_q=\lambda\,\left[\left(\frac{(1-cy)\sin^2\theta}{6}+\Omega(y)\cos^2\theta\right)m_1^2+w(y)m_2^2\,+\right.\\
\left.\Omega(y)m_3^2-2w(y)f(y)\cos\theta m_1m_2-2\Omega(y)\cos\theta m_1m_3+2w(y)f(y)m_2m_3\right].
\end{multline}

For pulsating strings in $L^{p,q,r}$ one can make a similar analysis. The metric can be written in the form
\eq{
ds^2_{L^{p,q,r}}=\frac{\rho^2}{4\Delta_x}dx^2+\frac{\rho^2}{4\Delta_y}dy^2+\hat{G}_{pq}dy^p dy^q,\label{metric_Lpqr}
}
where
\eq{
\begin{split}
\Delta_x&=x(\alpha-x)(\beta-x)-\mu=(x-x_1)(x-x_2)(x-x_3),\,0<x_1<x_2<x_3,\\
\Delta_y&=(1-y^2)(\alpha+\beta+(\alpha-\beta)y)/2,\,\rho^2=(\alpha+\beta+(\alpha-\beta)y)/2-x.
\end{split}
}
The coordinates $x$ and $y$ have the ranges $x_1\leq x\leq x_2$ and $-1\leq y\leq 1$. In \eqref{metric_Lpqr} $\hat{G}_{pq}=\hat{G}_{pq}(x,\,y)$ is the remaining part of metric associated with $\phi,\,\chi,\,\psi$ coordinates, denoted here as $p,\,q=1,\,2,\,3,\,\,\,\,y^1=\phi,\,\,y^2=\chi,\,\,y^3=\psi$
\begin{multline}
\hat{G}_{pq}dy^p dy^q=\\
=\left(\frac{(\alpha-x)^2(1-y)^2}{4\alpha^2}+\frac{\Delta_x(1-y)^2}{2\alpha^2\rho^2}+\frac{(\alpha-x)^2\Delta_y}{4\alpha^2\rho^2}\right)d\phi^2+d\chi^2+
\left(\frac{(\beta-x)^2(1+y)^2}{4\beta^2}\,+\right.\\
\left.\frac{\Delta_x(1+y)^2}{2\beta^2\rho^2}+\frac{(\beta-x)^2\Delta_y}{4\beta^2\rho^2}\right)d\psi^2+
\frac{(\alpha-x)(1-y)}{\alpha}d\phi d\chi+\left(\frac{(\alpha-x)(\beta-x)(1-y^2)}{2\alpha\beta}\,+\right.\\
\left.\frac{\Delta_x(1-y^2)}{\alpha\beta\rho^2}-\frac{(\alpha-x)(\beta-x)\Delta_y}{2\alpha\beta\rho^2}\right)d\phi d\psi+\frac{(\beta-x)(1+y)}{\beta}d\chi d\psi.
\end{multline}
The pulsating string ansatz is
\begin{align}
x^1&=x^1(\tau)=x(\tau),\,\,\,\, &x^2=x^2(\tau)=y(\tau),&\,\,\,\,\\
y^1&=\phi=m_1\sigma+h^1(\tau),&y^2=\chi=m_2\sigma+h^2(\tau),\quad\,\,\,\,&y^3=\psi=m_3\sigma+h^3(\tau).
\end{align}
The potential has the form
\begin{multline}
V(x,y)=\lambda\,\hat{G}_{pq}(x,\,y)m_p m_q=\\
=\lambda\,\left[\left(\frac{(\alpha-x)^2(1-y)^2}{4\alpha^2}+\frac{\Delta_x(1-y)^2}{2\alpha^2\rho^2}+
\frac{(\alpha-x)^2\Delta_y}{4\alpha^2\rho^2}\right)m_1^2+m_2^2+\left(\frac{(\beta-x)^2(1+y)^2}{4\beta^2}\,+\right.\right.\\
\left.\frac{\Delta_x(1+y)^2}{2\beta^2\rho^2}+\frac{(\beta-x)^2\Delta_y}{4\beta^2\rho^2}\right)m_3^2+
\frac{(\alpha-x)(1-y)}{\alpha}m_1m_2+\left(\frac{(\alpha-x)(\beta-x)(1-y^2)}{2\alpha\beta}\,+\right.\\
\left.\left.\frac{\Delta_x(1-y^2)}{\alpha\beta\rho^2}-\frac{(\alpha-x)(\beta-x)\Delta_y}{2\alpha\beta\rho^2}\right)m_1m_3+\frac{(\beta-x)(1+y)}{\beta}m_2m_3\right].
\end{multline}

The above perturbations to the free actions will produce the corrections to the energy and therefore to the anomalous dimensions. In order to calculate the corrections to the energy as perturbations due to the above potentials, however, we need the normalized wave functions associated to $Y^{p,q}$ and $L^{p,q,r}$ spaces and all these issues are subject to the next section.


\section{The Laplace-Beltrami operator and wave functions for $Y^{p,q}$ and $L^{p,q,r}$}
In this section we will compute the Laplace-Beltrami operator on $Y^{p,q}$ and derive the wave functions. The $L^{p,q,r}$ case is treated in \cite{Oota:2005}. We will show how the particular case of $T^{1,1}$ arises from $Y^{p,q}$.

The line element of $Y^{p,q}$ in global coordinates is explicitly given by \eqref{metric_Ypq}. Using the standard definition of the Laplace-Beltrami operator, we find
\eq{
\bigtriangleup_{Y^{p,q}}=\frac{6\partial_{\theta}\left(\sin\theta\partial_{\theta}\right)}{(1-cy)\sin\theta}+
\frac{\partial_y\left[(1-cy)w(y)q(y)\partial_y\right]}{1-cy}+\hat{G}_{pq}^{-1}\partial_p\partial_q.\label{laplas}
}
The Schr\"{o}dinger equation for the wave function is
\begin{equation}
\bigtriangleup_{Y^{p,q}}\,\Psi(\theta,\,y,
\,\phi,\,\alpha,\,\psi)=-E^2\,\Psi(\theta,\,y,
\,\phi,\,\alpha,\,\psi).\label{schro}
\end{equation}
To separate the variables, we define $\Psi$ as
\begin{equation}
\Psi(\theta,\,y,
\,\phi,\,\alpha,\,\psi)= f_1(\theta)f_2(y)\exp(il_1\phi)\exp(il_2\alpha)\exp(il_3\psi),\,\,\,\,l_1,\,l_2,\,l_3\in\mathbb{Z}.\label{wave}
\end{equation}
With this choice we can solve for the eigenfunctions replacing the derivatives along Killing directions $(\partial_{\phi},\,\partial_{\alpha},\,\partial_\psi)$ by $(il_1,\,il_2,\,il_3)$ correspondingly. Writing \eqref{wave} in \eqref{schro}, and since \eqref{laplas}, we arrive at the following equation for $f_1(\theta)$
\begin{equation}
\left[\dfrac{1}{\sin\theta}\dfrac{d}{d\theta}\left(\sin\theta\dfrac{d}{d\theta}\right)
- \dfrac{1}{\sin^2\theta}\left(l_1+\cos\theta l_3\right)^2 \right]\,f_1(\theta)=-E_1^2\,f_1(\theta).
\end{equation}
It is convenient to define a new variable $z_1=\cos\theta$. Then, the equation can be written as
\begin{equation}
\left\lbrace
(1-z_1^2)\,\dfrac{d^2}{dz_1^2}-2z_1\,\dfrac{d}{dz_1}
-\dfrac{1}{1-z_1^2}\left(l_1+z_1\,l_3\right)^2+E_1^2\right\rbrace\,f_1(z_1)=0.
\end{equation}
This hypergeometric equation has already been encountered in \cite{Arnaudov:2010}, and its normalized solution (wave-function) is given there.

For $f_2(y)$ we have
\eq{
\begin{split}
\left[\frac{d^2}{dy^2}-\frac{6y(1-cy)}{a-3y^2+2cy^3}\frac{d}{dy}+\frac{1}{2(a-3y^2+2cy^3)}\left(E^2(1-cy)-6E_1^2\,-\right.\right.\\
\left.\left.\frac{9(1-cy)}{q(y)}(l_3-l_2f(y))^2-\frac{l_2^2(1-cy)}{w(y)}\right)\right]f_2(y)=0.
\end{split}
\label{eqf2}
}
This is an equation with four regular singular points, so it can be converted to the standard form of the Heun's equation~\cite{Ronveaux} by the following ansatz
\eq{
f_2(y)=(y-y_1)^{\mu_1}(y-y_2)^{\mu_2}(y-y_3)^{\mu_3}g(\tilde{y}),\,\,\tilde{y}=\frac{y-y_1}{y_2-y_1},
}
where $y_i,\,i=1,2,3,$ are the roots of $q(y)$, and
\eq{
\begin{split}
\mu_1^2&=\left[(4+16cy_1+5ac^2+9c^2y_1^2+2ac^3y_1)l_2^2-(48y_1+36c(a+y_1^2)+24ac^2y_1)l_2l_3\,+\right.\\
&\left.(36(a+y_1^2)+72acy_1)l_3^2\right]/\left[64c^2(y_1^2+2y_2y_3)^2\right],\\
\mu_2^2&=\left[(4+16cy_2+5ac^2+9c^2y_2^2+2ac^3y_2)l_2^2-(48y_2+36c(a+y_2^2)+24ac^2y_2)l_2l_3\,+\right.\\
&\left.(36(a+y_2^2)+72acy_2)l_3^2\right]/\left[64c^2(y_2^2+2y_1y_3)^2\right],\\
\mu_3^2&=\left[(4+16cy_3+5ac^2+9c^2y_3^2+2ac^3y_3)l_2^2-(48y_3+36c(a+y_3^2)+24ac^2y_3)l_2l_3\,+\right.\\
&\left.(36(a+y_3^2)+72acy_3)l_3^2\right]/\left[64c^2(y_3^2+2y_1y_2)^2\right].
\end{split}
}
Thus, $g(\tilde{y})$ satisfies the canonical Heun's equation
\eq{
\frac{d^2g}{d\tilde{y}^2}+\left(\frac{\gamma}{\tilde{y}}+\frac{\delta}{\tilde{y}-1}+\frac{\epsilon}{\tilde{y}-\hat{a}}\right)\frac{dg}{d\tilde{y}}+
\frac{\hat{\alpha}\hat{\beta}\tilde{y}-k}{\tilde{y}(\tilde{y}-1)(\tilde{y}-\hat{a})}g=0,
}
where
\begin{multline}
\hat{a}=\frac{y_3-y_1}{y_2-y_1},\,\gamma=2\mu_1+1,\,\delta=2\mu_2+1,\,\epsilon=2\mu_3+1,\,\hat{\alpha}+\hat{\beta}=\gamma+\delta+\epsilon-1,\\
\hat{\alpha}\hat{\beta}=\frac{1}{y_2-y_1}\left(\mu_1^2+\mu_2^2+\mu_3^2+2(\mu_1\mu_2+\mu_1\mu_3+\mu_2\mu_3)+2(\mu_1+\mu_2+\mu_3)-\frac{E^2}{4}\right),\\
k=\frac{1}{y_2-y_1}\left(2\mu_1^2(y_1-\frac{3}{4c})+2\mu_2^2(y_2-\frac{3}{4c})+2\mu_3^2(y_3-\frac{3}{4c})-y_1(2\mu_2\mu_3+\mu_2+\mu_3)\,-\right.\\
\left.y_2(2\mu_1\mu_3+\mu_1+\mu_3)-y_3(2\mu_1\mu_2+\mu_1+\mu_2)+\frac{E^2-6E_1^2}{4c}+\frac{3c^2l_2^2-12cl_2l_3-36l_3^2}{32c}\right).
\end{multline}
The solution to the canonical Heun's equation is given by Heun functions \cite{Ronveaux}. However, we have to ensure that the solution $f_2(y)$ is square integrable to the measure for $y$. Thus, we make the following restriction on the parameter
$\hat{\alpha}=-n,\,\,\,\,n \in\mathbb{N}$. Then, the solution and wave-function can be given in terms of Heun polynomials \cite{Ronveaux}. But since they cannot be integrated analytically, we are not able to calculate the correction to the energy due to the potential $V$.

For the case of $L^{p,q,r}$ after similar calculations one arrives at two Heun's equations~\cite{Oota:2005}, which makes the problem of finding the correction to the energy even more difficult.

A special case of $Y^{p,q}$ is $T^{1,1}$ ($a\mapsto3,\,c\mapsto0;\,y_1\mapsto-1,\,y_2\mapsto1,\,y_3\mapsto\infty$), so that \cite{Arnaudov:2010}
\eq{
\theta\mapsto\theta_1,\,y\mapsto z_2=\cos\theta_2,\,\phi\mapsto\phi_1,\,\alpha\mapsto\frac{\phi_2}{6}\Longrightarrow l_2\mapsto 6l_2,\,\psi\mapsto\psi.
}
The measure and the potential $V$ for $Y^{p,q}$ are easily reduced to the ones in \cite{Arnaudov:2010}. Also, the equation for $f_1(z_1)$ already has the same form as in \cite{Arnaudov:2010}. Equation \eqref{eqf2} transforms to a hypergeometric equation ($E^2=6E_1^2+6E_2^2+9l_3^2$)
\begin{equation}
\left\lbrace
(1-z_2^2)\,\dfrac{d^2}{dz_2^2}-2z_2\,\dfrac{d}{dz_2}
-\dfrac{1}{1-z_2^2}\left(l_2+z_2\,l_3\right)^2+E_2^2\right\rbrace\,f_2(z_2)=0,
\end{equation}
which coincides with the corresponding equation in \cite{Arnaudov:2010}. Thus, we have derived completely the $T^{1,1}$ case from our analysis of $Y^{p,q}$.

\section{Conclusion}
Our study was motivated by the recently suggested duality between string theory in
$AdS_5\times T^{1,1}$ and ${\cal N}=1$ superconformal field theory. The results obtained so far
provide important understanding of the string/gauge theory dualities and in particular the region of strong coupling
\cite{Klebanov:1998hh}-\cite{Giataganas:2009}. The purpose of
this paper was to look for pulsating string solutions in $AdS_5\times Y^{p,q}$ and $AdS_5\times L^{p,q,r}$ backgrounds. The class of pulsating strings has been used to study AdS/CFT correspondence in the case of $AdS_5\times S^5$ \cite{Minahan:2002rc,Dimov:2004xi,Smedback:1998yn},
and the corrections to the string energy have been associated with anomalous dimensions of certain type of operators.

Here we consider a generalized string ansatz for pulsating strings in the $Y^{p,q}$ and $L^{p,q,r}$ parts of the geometry.
Next, we derive the corrections to the classical energy. From AdS/CFT point of
view the corrections give the anomalous dimensions of the operators in SYM theory and therefore they are of primary interest. After that, we consider the Laplace-Beltrami operator on $Y^{p,q}$ and partially find the wave functions. Because Heun polynomials are impossible to be integrated analytically, we are not able to quantize the resulting theories semi-classically and to obtain the corrections to the energy. Consequently, we note as a final comment that it is of great interest to perform a numerical analysis of this problem.


\begin{theacknowledgments}
This work was supported in part by the Austrian Research Funds FWF P22000 and I192,
NSFB VU-F-201/06 and DO 02-257.
\end{theacknowledgments}

\bibliographystyle{aipproc}

\end{document}